\documentclass[prb,superscriptaddress,showpacs,
               reprint]{revtex4-1}

\usepackage[utf8]{inputenc}
\usepackage{amsmath,amsfonts,amssymb}
\usepackage{pifont}
\usepackage{rotating}
\usepackage{graphicx}    
\usepackage{epstopdf}
\usepackage{color}
\usepackage[caption=false]{subfig}
\usepackage{soul}

\usepackage{hyperref}
\definecolor{dark-red}{rgb}{0.9,0.15,0.15}
\definecolor{dark-blue}{rgb}{0.15,0.15,0.4}
\definecolor{medium-blue}{rgb}{0,0,0.5}
\hypersetup{
    colorlinks, linkcolor={black},
    citecolor={dark-blue}, urlcolor={medium-blue}
}

\begin{document}

\title{ The spin-1/2 coupled tetramer system Ba(TiO)Cu$_4$(PO$_4$)$ _4$ probed by magnetization, specific heat, and $^{31}$P-NMR  }

\author{Vinod Kumar}
\email{vinodkiitb@gmail.com}
\affiliation{Department of Physics, Indian Institute of Technology Bombay, Mumbai 400076, India}

\author{Aga Shahee}
\affiliation{Department of Physics, Indian Institute of Technology Bombay, Mumbai 400076, India}
\affiliation{Center for Novel States of Complex Materials Research, Department of Physics and Astronomy, Seoul National University, Seoul 151-747, Republic of Korea}

\author{S. Kundu}
\affiliation{Department of Physics, Indian Institute of Technology Bombay, Mumbai 400076, India}

\author{M. Baenitz}
\affiliation{Max Planck Institut für Chemische Physik fester Stoffe, Nöthnitzer Strasse 40, 01187 Dresden, Germany}

\author{A.V. Mahajan}
\email{mahajan@phy.iitb.ac.in}
\affiliation{Department of Physics, Indian Institute of Technology Bombay, Mumbai 400076, India}


\begin{abstract}
	
We present the synthesis and a detailed investigation of structural and magnetic properties of polycrystalline  Ba(TiO)Cu$_{4}$(PO$_{4}$)$_{4}$ (BTCPO) via x-ray diffraction, magnetic susceptibility, heat capacity, and $^{31}$P Nuclear Magnetic Resonance (NMR) measurements. BTCPO has a 2D layered structure with interlinked Cu$_{4}$O$_{12}$  tetramer units. A broad maximum is observed around 16.5 K in our magnetization data accompanied by a sharp anomaly around $T$ = 9.5 K in the heat capacity. An anomaly at about $T$ = 10 K is also found in the temperature dependence of the $^{31}$P NMR spin-lattice relaxation rate $1/T_1$. A power law behavior for the heat capacity as well as for the $^{31}$P $1/T_1$ below the ordering temperature could be obtained. The $^{31}$P NMR lineshape is asymmetric and the NMR shift tracks the bulk spin-susceptibility. We estimated the isotropic and axial components of the hyperfine coupling tensor to be as the $A^{iso}_{hf} \backsimeq 6794$  $\rm{Oe/\mu_{B}}$ and $A^{ax}_{hf} \backsimeq 818$  $\rm{Oe/\mu_{B}}$, respectively.

\end{abstract}


\date{\today}

\maketitle

\section{Introduction}
Low-dimensional quantum spin systems have been the harbinger of high-$T_{\rm{c}}$ superconductivity in the cuprates. On their own as well, they evidence
exotic magnetic ground states and excitations. This has lead to an enduring
interest in quantum magnets and efforts continue to be made to obtain
a better understanding of the diversity of the underlying physics in
actual experimental realizations.

Ba(TiO)Cu$_{4}$(PO$_{4}$)$_{4}$ (BTCPO) has a tetragonal structure
in which Cu$_{4}$O$_{12}$ assemblies are formed by four corner sharing square
planar CuO$_{4}$ units. Each cluster of four CuO$_{4}$ units buckles from its planar orientation and gives rise to cupolas (see Fig.\ref{fig:1}). These Cu cupolas
interact with each other through PO$_{4}$ pyramids. In the recent past, interesting work has been reported on the system
BTCPO where quadrupolar magnetic
moments arise due to buckling of the plaquettes made up of four corner-sharing
square-planar CuO$_{4}$ units which give rise to magnetoelectric
activity\cite{kimura_nature}. The magnetic
susceptibility of BTCPO exhibits a broad maximum ($\sim$ 17 K) possibly due
to short range order in the layers or possibly within the cupola. Finally, three-dimensional (3D) long range order (LRO) sets in at
9.5 K due to interlayer interactions\cite{kimura_nature}.

Detailed work pertaining to magnetism and magnetoelectricity
of BTCPO (and on Sr as well as on Pb analogs) as also neutron diffraction and dielectric measurements have been reported\cite{kimuraInorgchem,babkevich2017magnetic,kimura2018cation}. Further, magnetization, heat capacity, $^{31}$P NMR and density functional theory calculations have been reported in Sr(TiO)Cu$_{4}$(PO$_{4}$)$_{4}$\cite{SS_Islam_SCTPO}(STCPO). STCPO was found to order antiferromagnetically at a lower temperature of $ T\rm{_N}=6.2$ K and a spin gap of $ \Delta/k\rm{_B} = 10.6$ K was found. Recently in Ref.\cite{Y.Kato_2019_PRB} a combined experimental and theoretical magnetoelectric properties of STCPO, BTCPO and Pb(TiO)Cu$_{4}$(PO$_{4}$)$_{4}$ (PTCPO) has been reported in detail.          

In this paper, we present the structural and magnetic properties of
BTCPO together with heat capacity and $^{31}$P Nuclear Magnetic
Resonance (NMR) measurements. We find a broad peak in the susceptibility ($\chi(T)$) around 16.5 K. No sharp anomaly is seen in $\chi(T)$ below 16.5 K which is in strong contrast to the heat capacity which shows a clear $\rm\lambda$-like anomaly characteristic of three-dimensional (3D) long-range order (LRO). The $^{31}$P NMR shift ($K$) shows a broad maximum at $T\sim$ 15 K. We obtain power-law dependencies for the heat capacity and the $^{31}$P NMR $1/T_1$ below the ordering temperature.   
                   
\section{Experimental Details}
A polycrystalline sample of BTCPO was prepared by conventional solid state reaction method using high purity initial materials: BaCO$ _3 $ (Alfa Aesar 99.95\% ), TiO$ _2 $ (Alfa Aesar 99.9\%), NH$ _4 $H$ _2 $PO$ _4 $ (Loba Chemie ultra pure) and CuO (Alfa Aesar 99.995\%). Powder x-ray diffraction (XRD) measurements were performed at room temperature with Cu ${K _\alpha }$ radiation ($\lambda= 1.54182$ \AA) on a PANalytical X'PertPRO diffractometer. Magnetization measurements were carried out in the temperature range 2-400 K and the field range 0-70 kOe using a Quantum Design SVSM. Heat capacity measurements were done in the temperature range 2-295 K in the field range 0-90 kOe using the heat capacity option of a Quantum Design PPMS. $^{31}$P NMR measurements have been done using pulsed NMR techniques in a fixed magnetic field of 93.954 kOe  with a Tecmag spectrometer. The temperature was varied between 4-300 K using an Oxford continuous flow cryostat.

\section{Results and discussion}  
 In this section, we present the results of our XRD, magnetization, heat capacity and $^{31}$P NMR measurements followed by a discussion of the obtained data. 
 \subsection{Crystal structure}
\begin{figure}[h!]
\centering
\includegraphics[width=8.5cm, height=6cm]{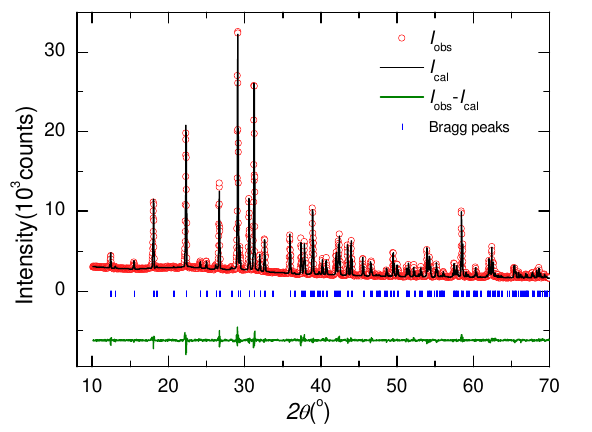}
\caption{(Color online) X-ray diffraction (XRD) data for BTCPO obtained at 300 K with a wavelength of 1.54182 $\rm{\AA}$. The red circles represent observed XRD pattern and the calculated XRD pattern (from Rietveld refinement) is shown by a solid line. The expected Bragg peak positions are shown as vertical blue bars and the difference between the observed and the calculated intensities is shown by green line.}
\label{fig:2}
\end{figure}

  The powder XRD pattern of BTCPO (see Fig.\ref{fig:2}) could be indexed within the space group $P$ 4 21 2 (space group no. 90) without any impurity peaks. The refinement was carried out with the FULLPROF SUITE\cite{CARVAJAL_fullprof}. The refinement parameters are shown in Table \ref{tab:refinement parameters}. The lattice parameters for BTCPO are $a$ = $b$ = 9.6084 \AA, $c$ = 7.1236 \AA. The refined atomic coordinates are shown in Table \ref{tab:atomic_position}. Our results are in agreement with those of Ref. \cite{kimuraInorgchem}.\\
 
 \begin{table}[h!]
 \centering
 \begin{tabular}{|c|c|}
 \hline Space group & $P$ 4 21 2 \\ 
 \hline Crystal system & Tetragonal \\ 
 \hline Lattice parameters & $a=b=$9.6084 \AA, $c$=7.1236 \AA \\ 
 \hline Refinement parameters & $R_{p}=14.1, R_{wp}=11.2 $\\
                              & $R_{exp}=6.99,gof=2.72$\\ 
 \hline 
 \end{tabular} 
 \caption{Lattice parameters and refinement parameters after crystal structure refinement of BTCPO by Fullprof.}
 \label{tab:refinement parameters}
 \end{table}
 
 \begin{table}[h!]
  \centering
    \begin{tabular}{|c|c|c|c|c|c|c|}
    \hline Atom & Wyk. Pos.  & x & y & z & Occ. & B($\rm\AA^2$) \\ 
    \hline Ba &2a  & 0.0000 & 0.0000 & 0.0000 & 1 & 1.6(2) \\ 
    \hline Cu & 8$g$ & 0.2662 & 0.9928 & 0.4007 & 1 &2.4(2) \\ 
    \hline Ti &2c  & 0.5000 & 0.0000 & 0.9717 & 1 &1.7(3)\\ 
    \hline P & 8$g$ & 0.2839 & 0.1930 & 0.7458 & 1 &2.7(4)\\ 
    \hline O1 & 8$g$ & 0.3558 & 0.1302 & 0.5581 & 1 &0.734298\\ 
    \hline O2 & 8$g$  & 0.1237 & 0.1474 & 0.7254 & 1 &0.884316\\ 
    \hline O3 & 8$g$ & 0.3315 & 0.1201 & 0.9256 & 1 &0.868525\\ 
    \hline O4 & 2c  & 0.5000 & 0.0000 & 0.2074 & 1 &0.947482\\ 
    \hline O5 & 8$g$  & 0.3056 & 0.3384 & 0.7635 & 1 &0.836942\\ 
    \hline 
    \end{tabular} 
    \caption{Atomic positions in BTCPO are presented after refinement of the room temperature XRD data by Fullprof. Wyckoff positions, occupancy and thermal parameters are shown for the individual atoms.}
    \label{tab:atomic_position}
    \end{table}
 
\begin{figure}[h!]
\centering
\includegraphics[width=8.5cm, height=12cm]{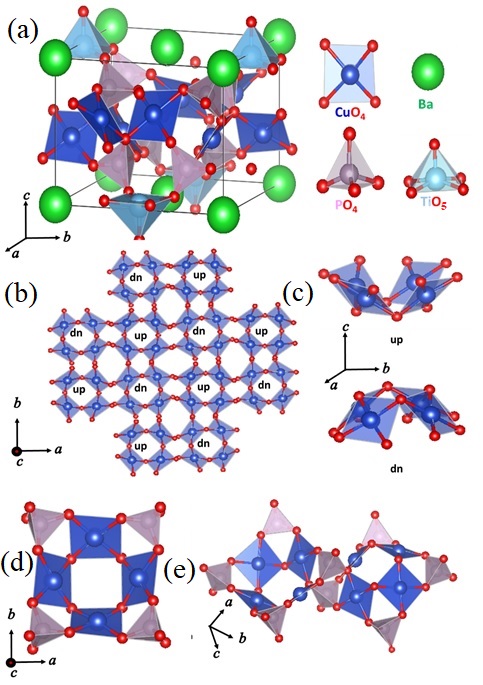}
\caption{(Color online) Schematic representation of the crystal structure of BTCPO in the space group $P$ 4 21 2. (a) An ideal unit cell of BTCPO is shown with CuO$_4$ plaquettes (blue), PO$_4$ tetrahedra (pink) and TiO$_5$ pyramids (cyan). (b) Cu$_4$O$_{12}$ cupola structure facing upward (up) and downward (dn) is shown in a-b plane. (c) Individual Cu$_4$O$_{12}$ cupola structure facing upward (up) and downward (dn) is shown. (d) Connectivity of P with Cu via O is presented within Cu$_4$O$_{12}$. Here Cu is hyperfine coupled with P through O. (e) Connectivity of different Cu$_4$O$_{12}$ assembly via PO$_{4}$ pyramids is shown.}
\label{fig:1}
\end{figure}

 The unit cell of BTCPO is shown in Fig.\ref{fig:1}(a) with CuO$_4$ plaquettes connected to TiO$_5$ pyramids through PO$_4$ tetrahedra. In Fig.\ref{fig:1}(b) the (buckled) upward and downward facing Cu$_{4}$O$_{12}$ square cupolas have been shown. Individual up facing and down facing Cu$_4$O$_{12}$ Cupola is shown in Fig.\ref{fig:1}(c). How a Cu is coupled with P via O is shown in Fig.\ref{fig:1}(d and e), where one Cu atom is connected with P through three different O atoms. The Cu-O bond length varies from  1.921 $\mathrm\AA$ to 2.011 $\mathrm\AA$  and the O-P bond length has values from 1.418 $\mathrm\AA$ to 1.621 $\mathrm\AA$. The bond angle Cu-O-P has values between 56.8$^{\circ}$ to 122.8$^{\circ}$.       
 \begin{table}[h!]
 	\centering
 	\begin{tabular}{|c|c|}
 		\hline Bond length (\AA) & Bond angle ($^\circ$) \\ 
 		\hline $\mathrm{Cu-O1} = 1.934$ &$\angle  \mathrm {Cu-O1-P} = 122.8$ \\ 
 		\hline $\mathrm {O1-P} = 1.621$ &$\angle  \mathrm {Cu-O2-P} = 120.6$ \\ 
 		\hline $\mathrm {Cu-O2} = 1.921$ &$\angle  \mathrm {Cu-O5-P} = 56.8$ \\ 
 		\hline $\mathrm {O2-P} = 1.607$ &  \\ 
 		\hline $\mathrm {Cu-O5} = 2.011$ &  \\ 
 		\hline $\mathrm {O5-P} = 1.418$ &  \\ 
 		\hline 
 	\end{tabular} 
 	\caption{Bond lengths and the corresponding bond angles for the interaction paths Cu-O1-P, Cu-O2-P and Cu-O5-P in the unit cell of BTCPO.}
 	\label{tab:distance and angle}
 \end{table}                                                                                                                                                                  

   \subsection{Magnetic Susceptibility}
\begin{figure}[h!]
\centering
\includegraphics[width=8cm, height=6cm]{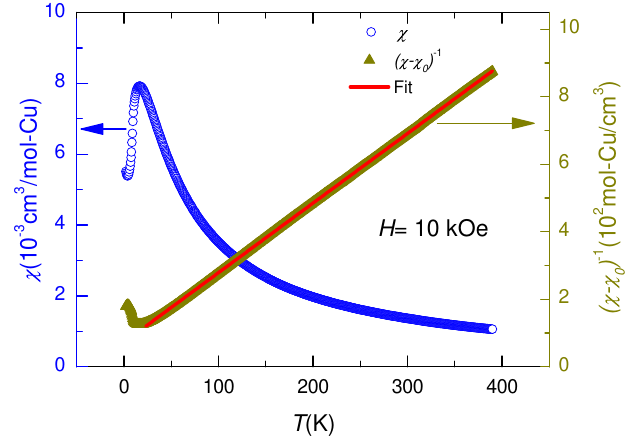}
\caption{(Color online) $T$ dependence of $\chi$, measured at $H$ = 10 kOe (blue open circles) and inverse susceptibility $1/(\chi-\chi_0) $, where $\chi_0$ is temperature independent susceptibility, as a function of $T$ (dark-yellow solid triangles). The linear Curie-Weiss fit is done in the $ T $-range 20 - 395 K (red solid line).}
\label{fig:3}
\end{figure}

  Fig.\ref{fig:3} shows the magnetization ($M$) divided by the magnetic field ($ H $) (${(\chi(T))=M(T)/H}$) for BTCPO as a function of temperature  along with the inverse susceptibility $ 1/(\chi-\chi_0)$. A fit to the high-temperature data (150-390 K) using the Curie-Weiss law ($ \chi = \chi\rm{_{0}} + C/(T-\theta_{CW} $)) yields a Curie-Weiss temperature $\theta\rm{_{CW}}= -35$ K which indicates antiferromagnetic interaction between the Cu$^{2+}$ ions. The $\theta\rm{_{CW}}$ value for BTCPO is higher than that of the isostructural STCPO compound which is 18.7 K. The temperature independent susceptibility $\chi\rm{_{0}}$ is obtained to be $-8.63\times 10^{-5} \rm{cm^{3}/mol-Cu} $. Furthermore, the effective magnetic moment is $\mu\rm{_{eff}} = 1.97$ $\rm{\mu_{B}/Cu}$ which is typical for Cu$^{2+}$ ion with a small spin-orbit coupling. On cooling, a broad maximum appears around 16.5 K in $\chi(T)$. The broad maximum is likely due to short range correlation within each cupola or within the two dimensional Cu-O layer.  
  
  We have further fitted the susceptibility, after subtracting the temperature independent susceptibility ($\chi-\chi_0$), to a uniform tetramer model with nearest neighbor interaction $J$ in the Heisenberg model 
  
  \begin{equation} \label{equ.1}     
       H =  -J \sum \boldsymbol{S_{k}}.\boldsymbol{S_{l}} 
    \end{equation}

    The expression for susceptibility \cite{Formula_Rectangular_Tetramer} ($\chi(T)$)  in this case is:
    
    \begin{equation} \label{eq:10}  
         \dfrac{\chi}{(g\mu_{B})^{2}}=\dfrac{2\beta(5+3e^{\beta J})}{(5+6e^{\beta J}+3e^{2\beta J}+2e^{2\beta J}cosh(i\beta J/2))}
        \end{equation}

        where $\beta = \dfrac{1}{k_{B}T}$, $k_{B}$ is Boltzmann constant, $\mu_{B}$ is the Bohr magneton, $i=\sqrt{(-1)}$ and $g$ is the electron $g$-factor.  
        
\begin{figure}[h!]
\centering
\includegraphics[width=8.5cm, height=6cm]{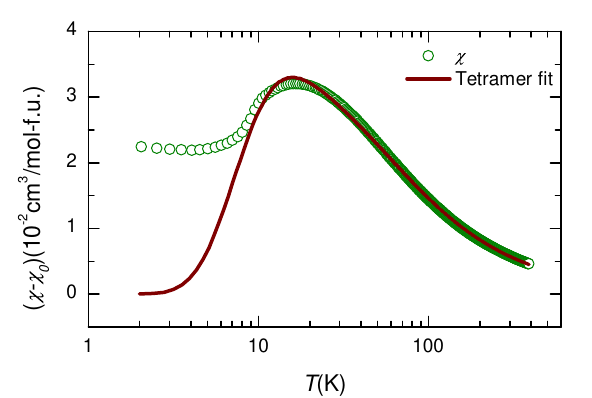}
\caption{(Color online) Magnetic susceptibility (after subtracting a $T$ independent susceptibility, $\chi_0$) data of BTCPO measured at $H$=10 kOe as a function of temperature (open green circles) together with a fit (solid wine line) to the uniform rectangular tetramer model ($J/k_{B}= 47$ K and $g=2.23$ ).}
\label{fig:4}
\end{figure}

Our fit (see Fig.\ref{fig:4}) yields $J/k_{B}\sim 47$ K and $g\sim2.23$. Although the simple tetramer model fits the $\chi(T)$ data well even a little below the broad maximum, non-zero intralayer and interlayer interactions among Cu$^{2+}$ ions were inferred from the analysis of the inelastic neutron scattering data\cite{kimura2018cation,kato2017_theory_fit}. Therefore a complete analysis which includes other intra-tetramer/cupola and inter-cupola interactions is needed.

   \subsection{Heat Capacity}
   To explore further the ground state and low energy excitations of BTCPO, we have measured the heat capacity $C\rm{_{P}}$$(T) $ as a function of temperature in various fields (0-90 kOe) as shown in Fig.\ref{fig:5}. Here it is clear that there is a sharp peak at $T\rm{_N}$= 9.5 K, which assures long-range ordering in the system. To inspect the behavior of $C_{p}$ below $T\rm{_N}$, variation of $C\rm{_{P}}$ against $T$ is shown in the inset of Fig.\ref{fig:5}. From this it becomes evident that in contrast to the Sr homologue $T_N$ of BTCPO is rather robust in a magnetic field and only a small shift of $T_N$ with field is observed. 
   
   The specific heat of a magnetic insulator comprises of the specific heat due to the lattice ($ C\rm{_{lattice}} $) and the  specific heat due to the magnetic entities ($ C\rm{_{m}} $)\cite{gopal2012specific}. 
   
   To estimate the lattice contribution ($ C_{lattice} $) in the absence of a suitable non-magnetic analog for BTCPO, a combination of Debye and Einstein terms, given below, have been used.
\begin{equation} \label{eq.11}
    C_{lattice}(T) = C_{debye}+C_{Einstein}
\end{equation}
where, $C_{Debye} = C_{D} [9k_{B} (\frac{T}{\theta_{D}})^3 \int_{0}^{\theta_{D}}{\frac{x^4e^x}{(e^x-1)^2}dx}]$ and\\

$C_{Einstein}=\sum{C_{Ei}} [3R (\frac{\theta_{Ei}}{T})^2 \dfrac{exp(\frac{\theta_{Ei}}{T})}{(exp(\frac{\theta_{Ei}}{T})-1)}]$.\\

One formula unit of Ba(TiO)Cu$_4$(PO$_4$)$ _4 $ has 27 atoms. Hence in each crystallographic direction there will be one acoustic and 26 optical modes of atomic vibrations\cite{ashcroft1976solid}. The Debye integral term in Eq.(\ref{eq.11}) represents three accoustic vibrational modes and the Einstein term in Eq.(\ref{eq.11}) accounts for 78 optical vibrational modes of phonons. This amounts to the condition $C_{D}=\frac{1}{27} $ and $\sum{C_{Ei}}=\frac{26}{27}$ or $ C_{D}+C_{Ei}=1 $ for each direction. The experimental   data were found to be fitted well (see Fig.{\ref{fig:5}) with the weightage factors corresponding to Debye model and Einstein model. Taking $n=27$, our fit yields $C_D=0.037$, $ \theta_D=150$ K, $C_{E1}=0.45$, $ \theta_{E1}=967$ K, $C_{E2}=0.36$, $ \theta_{E2}=393$ K, $C_{E3}=0.17$, $ \theta_{E3}=147$ K . Here it is worth to mention that this is a simple model to extract the basic characteristics of lattice dynamics which is complex in nature.

\begin{figure}[h!]
\centering
\includegraphics[width=8.5cm, height=5.5cm]{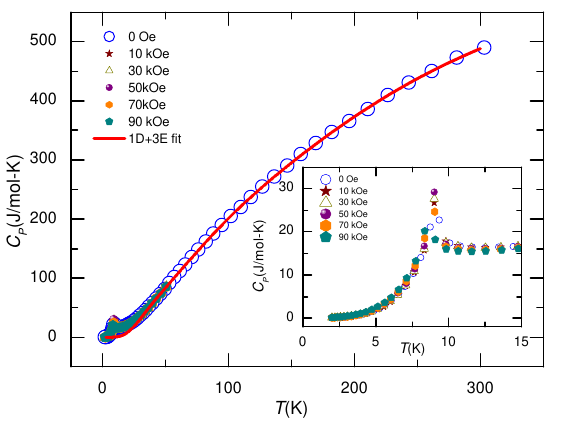}
\caption{(Color online) The $C\rm{_{P}}$($T$) data as a function of temperature at different magnetic fields is presented. The 1D+3E combination fit of $C\rm{_{P}}$($T$) data at 0 Oe is shown in solid red line. In the inset closer view of $C\rm{_{P}}$($T$) data where the system shows antiferromagnetic ordering has been shown ($\sim$ 9.5 K).}
\label{fig:5}
\end{figure}

The magnetic specific heat ($ C\rm{_{m}} $($T$)) has been extracted by subtracting $C\rm{_{lattice}}$ from $ C_{P}(T) $ data. In  Fig.\ref{fig:6}, $ C\rm{_{m}} $($T$) at different applied fields has been shown. One can notice that the peak position is nearly field dependent and the peak shifts by $\sim$1 K at 90 kOe . A power law ($ C\rm{_m}$ $\propto T^{\alpha} $) behavior is observed below $ T\rm{_N} $. From the Fig.\ref{fig:6} the extracted values of the power law exponent are 3.9 and 3.6 for zero field and 90 kOe respectively. The high power law exponent is due to 3D ordering\cite{joshua1998magnon} among Cu$^{2+}$ ions, which are interacting antiferromagnetically  and due to phonon-magnon coupling\cite{Pincuss-Winter}. A gap of 11.5 K has been seen in neutron scattering measurements\cite{kimura_nature}. An exponential decrease of magnetic heat capacity is expected for a system with an excitation gap. Given the limited temperature range of our $C\rm{_{P}}$$(T)$ data, it may not be easy to distinguish exponential behavior from a (large) power law behavior.   
 
\begin{figure}[h!]
\centering
\includegraphics[width=8.5cm, height=6cm]{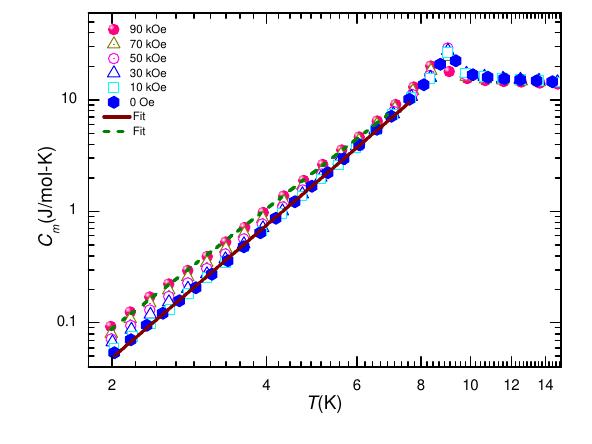}
\caption{(Color online) The $C_m$ data (log-log scale) for BTCPO at different applied magnetic fields have been shown as a function of temperature and the plot shows anomaly at $T_N=$ 9.5 K. Below ordering temperature $C_m$ is fitted with power law ($ C_m \propto T^{\alpha} $) as scales are in log-log, a linear fit is shown for 90 kOe and 0 Oe fields.}
\label{fig:6}
\end{figure}

After calculating $ C\rm{_{m}} $, the magnetic entropy change $ \Delta S\rm{_{m}} $ can be calculated simply by integrating $C_{m}/T$ with respect to temperature $T$. In Fig.\ref{fig:7} $ \Delta S\rm{_{m}} $ as a function of $T$ at different magnetic fields has been shown. It is notable that most of the entropy decrease has taken place above $T_N$ ($\simeq$ 9.5 K). The calculated $ \Delta S\rm{_{m}} $ in zero field was found to be $5.5$ (J/K-mol-Cu) upto 50 K, which is 95 \% of  $R$ln(2)$ = 5.762$ $\rm{(J/K-mol)} $, expected for a $ S=1/2 $ system. This clearly confirms the magnetic order in BTCPO is of long ranged nature. The apparent missing entropy probably is due to the uncertainty in the estimation of the phonon reference $ C\rm{_{lattice}} $.  

\begin{figure}[h!]
\centering
\includegraphics[width=8.5cm, height=6cm]{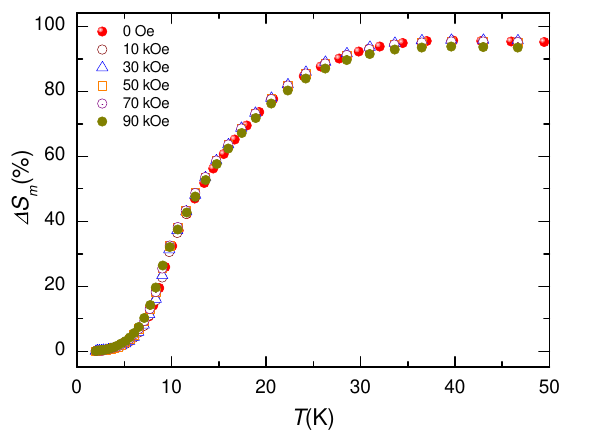}
\caption{(Color online) The magnetic entropy change ($\Delta S_m(T)$) as a function of temperature normalized to $R$ln2 for BTCPO in different magnetic fields is shown.}
\label{fig:7}
\end{figure}

    \subsection{$^{31}$P NMR}
    
    \subsubsection{NMR shift}
NMR is a robust local probe to study the static and dynamic properties of the magnetic systems. In BTCPO $^{31}$P nuclei are coupled to the magnetic spins of Cu$^{2+}$ through a transferred hyperfine coupling. As the $^{31}$P is a spin 1/2 nucleus, a single spectral line is expected for $^{31}$P NMR. Respective spectra of BTCPO at selected temperatures have been shown in Fig.\ref{fig:8}. We can see an asymmetric line shape which shifts with temperature. The asymmetric line shape is either due to an anisotropy of the spin susceptibility and/or anisotropy of the hyperfine coupling.
As shown in Fig.\ref{fig:8} the peak position of the spectrum shifts to higher frequencies as temperature is lowered, levels off around 15 K and then it shifts to lower frequencies with the further decrease of temperature down to 4 K.

\begin{figure}[h!]
\centering
\includegraphics[width=8cm, height=12cm]{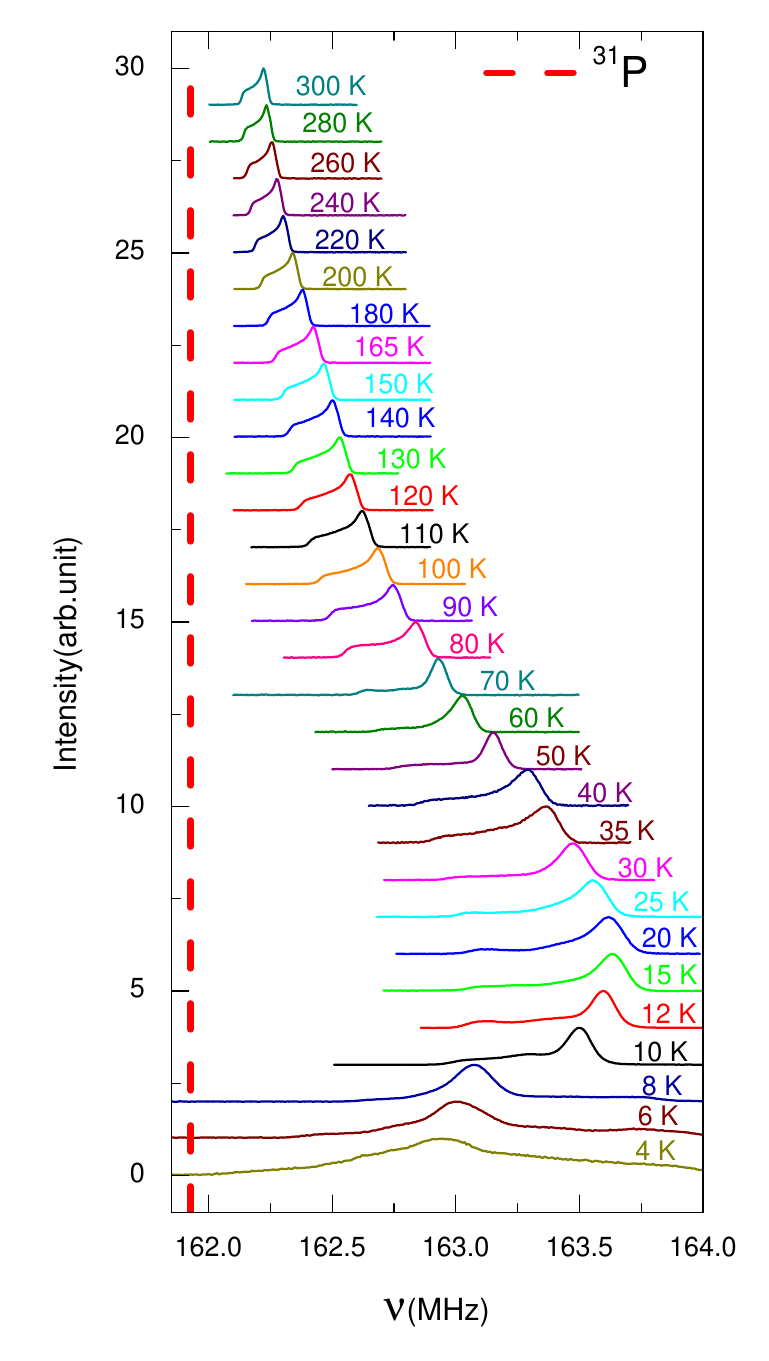}
\caption{(Color online) Temperature dependent $^{31}$P NMR spectra of BTCPO in a magnetic field of 93.954 kOe. The peak position shifts to higher frequencies as $T$ is lowered from 300 K to 15 K and then shifts to low frequencies on decreasing $T$ from 15 K to down to 4 K.}
\label{fig:8}
\end{figure}

In order to estimate the $^{31}$P NMR powder averaged line shifts ($K_{iso}$, $K_{aniso}$ and $K_{ax} $) from our experimentally taken spectra as a function of temperature, we compared the measured spectra with simulated lineshapes here. 
\begin{equation} \label{equ.2}
    K_{iso} = \dfrac{1}{3}(K_{X}+K_{Y}+K_{Z})
    \end{equation}
 
 \begin{equation} \label{equ.2}     
     K_{aniso} = \dfrac{1}{2}(K_{Y}-K_{X}) 
  \end{equation}
  
   \begin{equation} \label{equ.2}     
       K_{ax} = \dfrac{1}{6}(2K_{Z}-K_{X}-K_{Y}) 
    \end{equation}
    
    where $K_{X}$, $K_{Y}$ and $K_{Z}$ denote the principal components of the shift tensor $K$. 
    
    Fixing $K\rm{_{aniso}}=0$ and taking $K\rm{_{iso}}$, $K\rm{_{ax}}$ as variable parameters, we looked for good agreement between the simulated and experimental spectra. A few simulated spectra as a function of frequency  at selected  temperatures are shown in Fig.\ref{fig:9}. With $K\rm{_{aniso}}=0$, $K\rm{_{iso}}$ provides the average intrinsic  susceptibility, which is proportional to $ \dfrac{1}{3}(2K_{\perp}+K_{\parallel}) $ for an axially symmetric shift ($K_{X}$ = $K_{Y} = K_{\perp}$ and $K_{Z}=K_{\parallel}$). Here $K_{\parallel}$ means that the external field is parallel to the $Z$ direction and $K_{\perp}$ corresponds to the external field perpendicular to the $Z$ direction. To scrutinize the behavior of intrinsic susceptibility (static susceptibility) $K\rm{_{iso}}$ has been plotted as a function of temperature in Fig.\ref{fig:10} (left $y$-axis). The $K_{iso}(T)$ for the $^{31}$P site shows a broad maximum $\sim$ 15 K, which indicates short-range ordering in BTCPO and the right $y$-axis in Fig.\ref{fig:10} shows $\chi(T)$ for BTCPO which follows same as $K_{iso}(T)$. Here for BTCPO ordering temperature and maximum is at higher temperature comparative to STCPO.  We note that $K(T)$ reflects $\chi\rm{_{spin}}$$(T)$ explicitly and it can be written
 
 \begin{equation} \label{equ.2}
    K(T) = K_{0} + \dfrac{A_{hf}}{N_{A}\mu_B}\chi_{spin}
 \end{equation}
 Here, $K_{0}$ is the temperature independent chemical shift and $A_{hf}$ is the total hyperfine coupling between $^{31}$P nuclei and Cu$^{2+}$ spins. Fig.\ref{fig:10} shows the comparison of NMR-shift as a function of temperature with susceptibility as a function of temperature and it is clear that the shift tracks the susceptibility. In the inset of Fig.\ref{fig:10}, $K_{iso}$ is plotted as a function of $\chi$ by taking temperature ($T$) as an implicit parameter. The linear fit of the data for T $>$ 10, yields the isotropic part of hyperfine coupling $A^{iso}_{hf} \backsimeq 6794$  $\rm{Oe/\mu_{B}}$. In Fig.\ref{fig:11} the axial $^{31}$P NMR shift is shown as a function of temperature and it also has a broad maximum around 15 K. To extract the axial part of the hyperfine coupling linear fit of $K_{ax}$ vs $ \chi$ is shown in inset of Fig.\ref{fig:11} and $A^{ax}_{hf} \backsimeq 818$ $\rm{Oe/\mu_B}$. These values are comparable to the hyperfine coupling constants for isostructural STCPO \cite{SS_Islam_SCTPO}, which are 6539 $\rm{Oe/\mu_{B}}$ and 952 $\rm{Oe/\mu_{B}}$  for $A^{iso}_{hf}$ and $A^{ax}_{hf}$ respectively. It is noteworthy to mention that the NMR shift directly explore $\chi_{spin}$ and is free from impurity.        

\begin{figure}[h!]
\centering
\includegraphics[width=8cm, height=12cm]{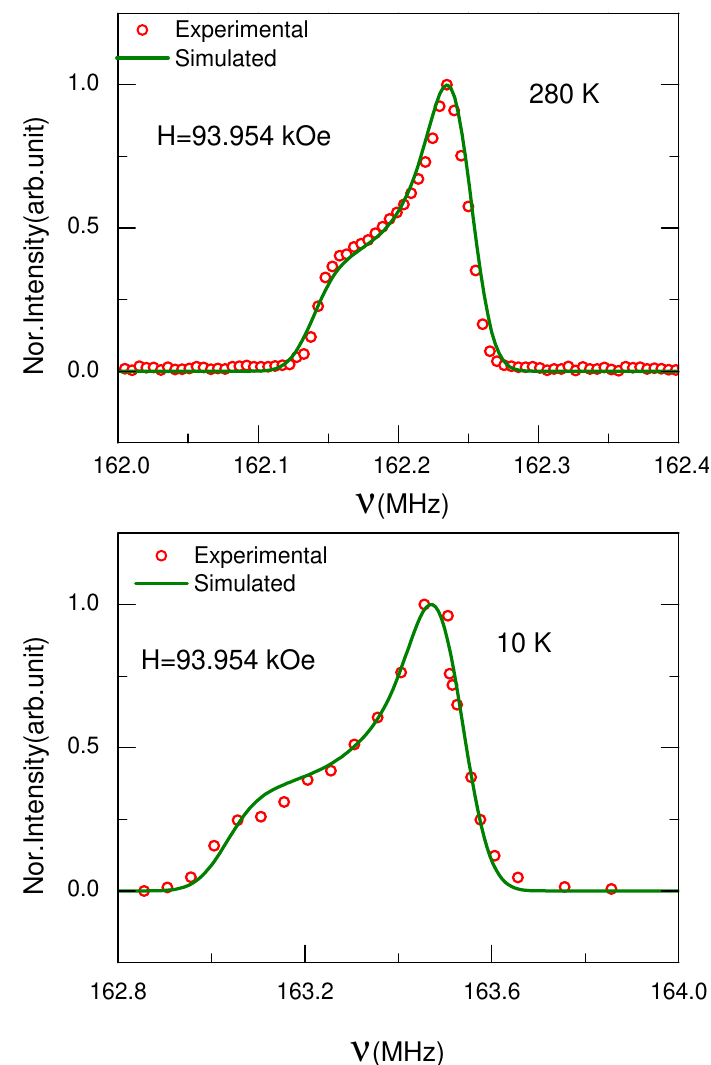}
\caption{(Color online) Simulated $^{31}$P NMR spectra and experimentally collected $ ^{31}$P NMR spectra at different temperatures are shown by green solid lines and red open circles for BTCPO respectively.}
\label{fig:9}
\end{figure}

\begin{figure}[h!]
\centering
\includegraphics[width=8.5cm, height=6cm]{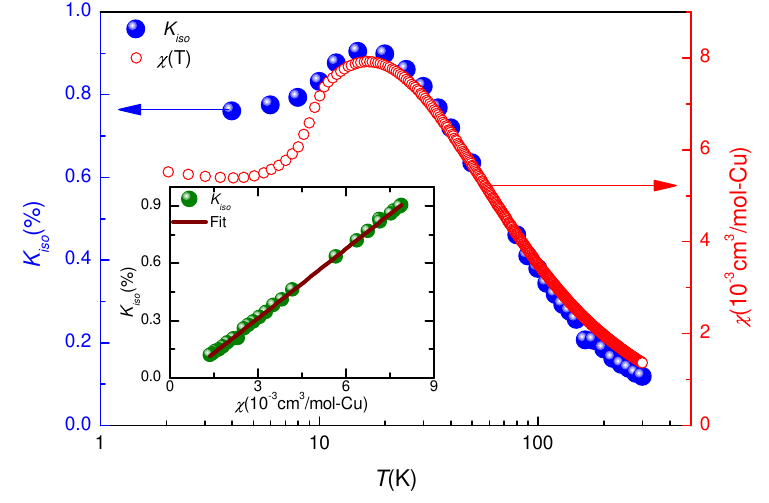}
\caption{(Color online) The left $y$-axis shows temperature dependent isotropic $^{31}$P NMR shift ($ K_{iso} $) as a function of temperature. The right $y$-axis shows temperature dependent magnetic susceptibility of BTCPO. In the inset variation of NMR-shift ($ K_{iso} $) vs $\chi$ is plotted with temperature as an implicit parameter and the solid line is the linear fit.}
\label{fig:10}
\end{figure}

\begin{figure}[h!]
\centering
\includegraphics[width=8.5cm, height=6cm]{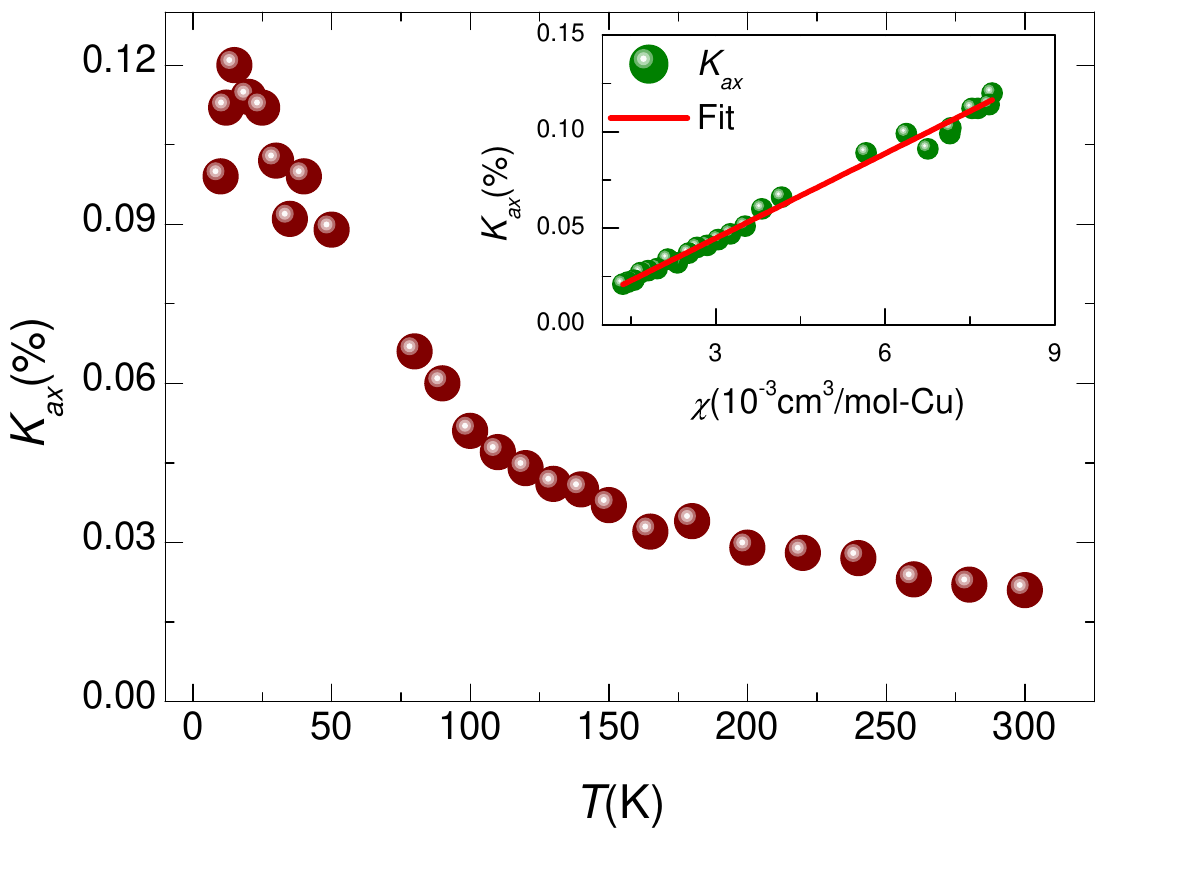}
\caption{(Color online) Temperature dependent axial $^{31}$P NMR-shift, $K_{ax}$ is shown as a function of temperature for BTCPO. It is showing broad maximum around 15 K. In the inset: $^{31}$P NMR shift vs $\chi$ is plotted for $K_{ax}$ with temperature as an implicit parameter and the solid line is the linear fit.}
\label{fig:11}
\end{figure}

\subsubsection{Spin-lattice relaxation rate, $1/T_{1}$}
The spin-lattice relaxation rate $1/T_1$ probes the low energy spin excitations\cite{moriya1956nuclear}. We have monitored the recovery of the longitudinal nuclear magnetization of $^{31}$P following a saturating pulse sequence. The recovery fits well to a single exponential function as expected for a $I=1/2$ nucleus.

\begin{figure}[h!]
\centering
\includegraphics[width=8.5cm, height=6cm]{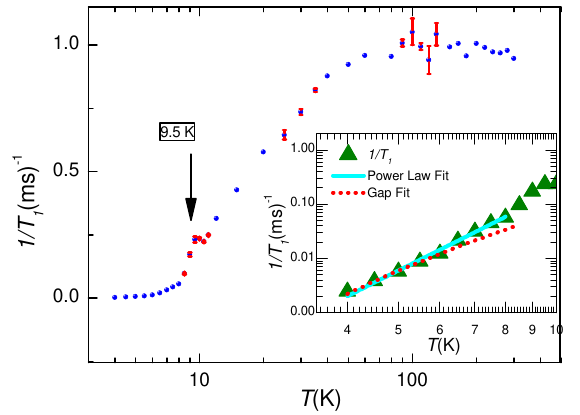}
\caption{(Color online) $1/T_{1}$ is plotted (in a semi-log scale) as a function of temperature with error bars at some selected points. In the inset: $1/T_1$ is plotted as a function of temperature in log-log scale. The fitting is shown for a power law ($1/T_1\propto T^{\alpha}$) in solid cyan line and for a gapped behavior ($1/T_1\propto T^2e^{(-\Delta/k_{B}T)}$) in dashed red line.}
\label{fig:12_2}
\end{figure}

\begin{figure}[h!]
\centering
\includegraphics[width=8.5cm, height=6cm]{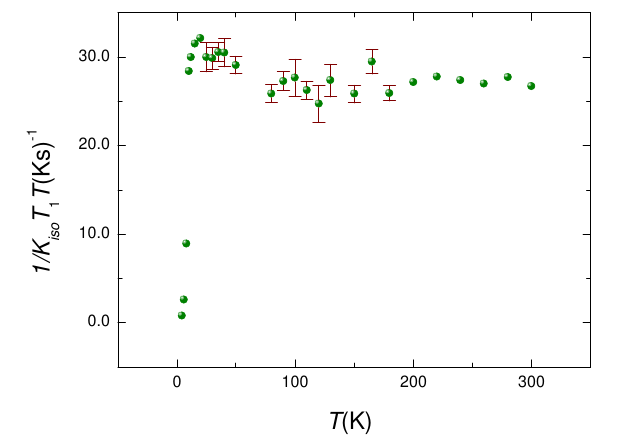}
\caption{(Color online) The $T$-dependence of $1/K_{iso}T_1T$ for BTCPO is shown here with error bars at some selected temperature.}
\label{fig:13_2}
\end{figure}

The $1/T_{1}$ values estimated from the single exponential function fit are plotted in Fig.\ref{fig:12_2} as a function of temperature. At high temperatures $(T>80 K)$, $1/T_{1}$ is temperature independent  and on lowering temperature down to 4 $K$ $1/T_{1}$ decreases rapidly towards zero.

In the paramagnetic state, the $1/T_1$ is expected to be $T$-independent as the fluctuation frequency is much greater than the NMR frequency. On lowering the temperature, $1/T_1$ decreases (below 80 K) for BTCPO and after going through an anomaly/plateau at about 10 K, it drops zero on further decreasing the temperature.  Generally $1/T_1$ is expected to rise on approach to LRO and exhibit a divergence near the ordering temperature but here no sharp divergence is seen which is likely due to filtering of antiferromagnetic fluctuation at $^{31}$P site which is symmetrically located between the Cu. So this anomaly at 9.5 K is due to LRO in BTCPO . 

The local moment contribution to the spin-lattice relaxation rate ($1/T_1$) is given by\cite{moriya1963,Mahajan_Li2VO4}
\begin{equation} \label{equ.8}
\frac{1}{T_1} = \dfrac{2k_{B}T}{\hbar^2}(\frac{\gamma_n}{\gamma_e})^2A_{hf}^2\Sigma_{\boldsymbol {q}}(\chi^{''}(\boldsymbol {q},\omega)/\omega)
\end{equation} 
where the sum is over the wave vector $\boldsymbol{q}$, $\gamma_{n}$ is the nuclear gyromagnetic ratio, $\gamma_{e}$ is the electron gyromagnetic ratio, $A_{hf}$ is assumed $\boldsymbol {q}$ independent hyperfine coupling constant, $\chi^{''}(\boldsymbol {q},\omega)$ is the $\boldsymbol {q}$ dependent imaginary part of the dynamical susceptibility at the nuclear Larmor frequency $\omega$ and the summation is over all wave vectors. In the limit $\omega\longrightarrow 0$ the summation is given by $\frac{\chi^{loc}(T)}{\omega_e}$, where $\chi^{loc}(T)$ is the static local moment susceptibility per mole per atom and $\omega_e$ is electron fluctuation frequency. In the high-$T$ limit the susceptibility is Curie-like ($C/T$) and the relaxation rate becomes $T$-independent and is $\frac{1}{T_1}=4\times\frac{2k_{B}}{\hbar^2}(\frac{\gamma_n}{\gamma_e})^2A_{hf}^2\frac{C}{N_A\omega_e}$, where the electron fluctuation frequency is $\omega_e=\frac{J}{\hbar}\sqrt{(\frac{8zS(S+1)}{3})}\simeq1.3\times10^{13}$rad/s where we assume uncorrelated fluctuations of 4 Cu neighbors of P. The number of nearest neighbors $z$ is taken as 4 and the exchange coupling $J/k_B$ is obtained (in a mean field approach) to be about 35 K from the $\theta\rm{_{CW}}$ value. Using the hyperfine field value $A\rm{^{iso}_{hf}}=$1698.5 Oe/$\mu\rm{_B}$ per Cu$^{2+}$ (1/4 of total hyperfine coupling constant, $A\rm{^{iso}_{hf}}=$6794 Oe/$\mu\rm{_B}$) in Eq. \ref{equ.8} , we obtain $1/T_1 \simeq 0.07$ ms$^{-1}$. This is around one order of magnitude smaller than the experimentally measured rate of about 1 ms$^{-1}$. A similar calculation of $^{31}$P NMR $1/T_1$ value in the paramagnetic region of STCPO\cite{SS_Islam_SCTPO} yields 0.12 ms$^{-1}$. This is also around one order of magnitude smaller than the experimentally measured rate of about 2 ms$^{-1}$. However the expected ratio of the spin-lattice relaxation rate ($1/T_1$) for BTCPO to STCPO is $\sim 0.6$ based on the Moria formula which is comparable to the experimentally inferred ratio of about 0.5. In Fig.\ref{fig:13_2}, 1/($K_{\rm {iso}} T_1 T)$ is plotted as a function of $T$. We see that 1/($K_{\rm {iso}} T_1 T)$ temperature independent above $\sim 10$ K. Based on Moriya formula, temperature independence of 1/($K_{\rm {iso}} T_1 T)$ would be expected when $q=0$ fluctuations dominate the relaxation (see Eq. \ref{equ.8}). The sharp decrease below 10 K is likely related to development of static moments below the ordering temperature. This dependence of 1/($K_{\rm {iso}} T_1 T)$ on $T$ is likely due to the development of long-range order below about 10 K. 

In the inset of Fig.\ref{fig:12_2}, $1/T_1$ is plotted as a function of $T$ and is seen to follow power law behavior ($1/T_1 \propto T^{\alpha}$) below $T_N=9.5$ K. The fit is seen to be good in the temperature range 8.5 - 4.0 K. The obtained exponent value is 4.9 and similar to that seen in our heat capacity data.       
        
 On the other hand, a fit to an activated behavior has been done for STCPO\cite{SS_Islam_SCTPO}, expecting that a two magnon (or Raman)\cite{moriya1956nuclear,N_Kaplan,D_Beeman} process might be responsible for nuclear spin-lattice relaxation as the spectral line shape is asymmetric. The equation used for such a fit is,

\begin{equation} \label{eq:12}
   1/T_{1} \propto T^2e^{-\Delta/k_{B}T}  
 \end{equation}
 
 where $\Delta$ is the gap. If we fit our data to Eq.\ref{eq:12} in the limited temperature range 6.0 - 4.0 K, we get $\Delta/k_{B}$  $\simeq$ 11 K. Note that in Ref.\cite{SS_Islam_SCTPO} as well, a fit to Eq.\ref{eq:12} has been performed in the limited temperature range 3.3 - 1.8 K to extract the spin-gap value. Overall, a power law dependence below the ordering temperature appears to be more robust.

  \section{Conclusion}
Using various experimental probes such as XRD, $ \chi (T) $, $ M(H) $, $ C_{P}(T) $ and NMR, properties of BTCPO  have been explored. Conventional rectangular tetramer model fitting for our magnetization data yields a value of the exchange coupling to be about 47 K. A clear anomaly is seen in our $C\rm{_{P}}$$(T)$ data suggestive of antiferromagnetic ordering around 9.5 K. The inferred entropy change is nearly equal to that expected from the spin entropy. $^{31}$P NMR lineshapes  are seen to be asymmetric. Analysis of the NMR data based on a hyperfine coupling anisotropy yields the isotropic and axial components of the hyperfine coupling tensor $A^{iso}_{hf} \backsimeq 6794$  $\rm{Oe/\mu_{B}}$, $A^{ax}_{hf} \backsimeq 818$  $\rm{Oe/\mu_{B}}$ respectively for BTCPO. The $^{31}$P NMR $1/T_1$ data exhibit an anomaly/plateau at about 10 K and analysis of the data below the ordering temperature suggests a power law variation with temperature. Given the large value of exponent ($\simeq 5$) so obtained, a simple exponential variation would also fit the data. $^{31}$P NMR $1/T_1$ follows a power law below ordering temperature with exponent value 6.8 for STCPO\cite{SS_Islam_SCTPO} which is somewhat larger than in BTCPO having exponent value 4.9. Exploration of the effect of dilution of the magnetic lattice on the magnetic and magnetoelectric properties appears to be an interesting direction for future pursuit.

\section{Acknowledgment}

We acknowledge the financial help and central facilities provided by IIT Bombay and IRCC. Vinod Kumar would like to acknowledge Dr. Koteswara Rao Bommisetti for sharing his views.


\bibliographystyle{apsrev4-1}
\bibliography{bib}

\begin{thebibliography}{18}%
\makeatletter
\providecommand \@ifxundefined [1]{%
 \@ifx{#1\undefined}
}%
\providecommand \@ifnum [1]{%
 \ifnum #1\expandafter \@firstoftwo
 \else \expandafter \@secondoftwo
 \fi
}%
\providecommand \@ifx [1]{%
 \ifx #1\expandafter \@firstoftwo
 \else \expandafter \@secondoftwo
 \fi
}%
\providecommand \natexlab [1]{#1}%
\providecommand \enquote  [1]{``#1''}%
\providecommand \bibnamefont  [1]{#1}%
\providecommand \bibfnamefont [1]{#1}%
\providecommand \citenamefont [1]{#1}%
\providecommand \href@noop [0]{\@secondoftwo}%
\providecommand \href [0]{\begingroup \@sanitize@url \@href}%
\providecommand \@href[1]{\@@startlink{#1}\@@href}%
\providecommand \@@href[1]{\endgroup#1\@@endlink}%
\providecommand \@sanitize@url [0]{\catcode `\\12\catcode `\$12\catcode
  `\&12\catcode `\#12\catcode `\^12\catcode `\_12\catcode `\%12\relax}%
\providecommand \@@startlink[1]{}%
\providecommand \@@endlink[0]{}%
\providecommand \url  [0]{\begingroup\@sanitize@url \@url }%
\providecommand \@url [1]{\endgroup\@href {#1}{\urlprefix }}%
\providecommand \urlprefix  [0]{URL }%
\providecommand \Eprint [0]{\href }%
\providecommand \doibase [0]{http://dx.doi.org/}%
\providecommand \selectlanguage [0]{\@gobble}%
\providecommand \bibinfo  [0]{\@secondoftwo}%
\providecommand \bibfield  [0]{\@secondoftwo}%
\providecommand \translation [1]{[#1]}%
\providecommand \BibitemOpen [0]{}%
\providecommand \bibitemStop [0]{}%
\providecommand \bibitemNoStop [0]{.\EOS\space}%
\providecommand \EOS [0]{\spacefactor3000\relax}%
\providecommand \BibitemShut  [1]{\csname bibitem#1\endcsname}%
\let\auto@bib@innerbib\@empty
\bibitem [{\citenamefont {Kimura}\ \emph
  {et~al.}(2016{\natexlab{a}})\citenamefont {Kimura}, \citenamefont
  {Babkevich}, \citenamefont {Sera}, \citenamefont {Toyoda}, \citenamefont
  {Yamauchi}, \citenamefont {Tucker}, \citenamefont {Martius}, \citenamefont
  {Fennell}, \citenamefont {Manuel}, \citenamefont {Khalyavin}, \citenamefont
  {Johnson}, \citenamefont {Nakano}, \citenamefont {Nozue}, \citenamefont
  {Rønnow},\ and\ \citenamefont {Kimura}}]{kimura_nature}%
  \BibitemOpen
  \bibfield  {author} {\bibinfo {author} {\bibfnamefont {K.}~\bibnamefont
  {Kimura}}, \bibinfo {author} {\bibfnamefont {P.}~\bibnamefont {Babkevich}},
  \bibinfo {author} {\bibfnamefont {M.}~\bibnamefont {Sera}}, \bibinfo {author}
  {\bibfnamefont {M.}~\bibnamefont {Toyoda}}, \bibinfo {author} {\bibfnamefont
  {K.}~\bibnamefont {Yamauchi}}, \bibinfo {author} {\bibfnamefont {G.~S.}\
  \bibnamefont {Tucker}}, \bibinfo {author} {\bibfnamefont {J.}~\bibnamefont
  {Martius}}, \bibinfo {author} {\bibfnamefont {T.}~\bibnamefont {Fennell}},
  \bibinfo {author} {\bibfnamefont {P.}~\bibnamefont {Manuel}}, \bibinfo
  {author} {\bibfnamefont {D.~D.}\ \bibnamefont {Khalyavin}}, \bibinfo {author}
  {\bibfnamefont {R.~D.}\ \bibnamefont {Johnson}}, \bibinfo {author}
  {\bibfnamefont {T.}~\bibnamefont {Nakano}}, \bibinfo {author} {\bibfnamefont
  {Y.}~\bibnamefont {Nozue}}, \bibinfo {author} {\bibfnamefont {H.~M.}\
  \bibnamefont {Rønnow}}, \ and\ \bibinfo {author} {\bibfnamefont
  {T.}~\bibnamefont {Kimura}},\ }\href {\doibase
  https://doi.org/10.1038/ncomms13039} {\bibfield  {journal} {\bibinfo
  {journal} {Nature communications}\ }\textbf {\bibinfo {volume} {7}},\
  \bibinfo {pages} {13039} (\bibinfo {year} {2016}{\natexlab{a}})}\BibitemShut
  {NoStop}%
\bibitem [{\citenamefont {Kimura}\ \emph
  {et~al.}(2016{\natexlab{b}})\citenamefont {Kimura}, \citenamefont {Sera},\
  and\ \citenamefont {Kimura}}]{kimuraInorgchem}%
  \BibitemOpen
  \bibfield  {author} {\bibinfo {author} {\bibfnamefont {K.}~\bibnamefont
  {Kimura}}, \bibinfo {author} {\bibfnamefont {M.}~\bibnamefont {Sera}}, \ and\
  \bibinfo {author} {\bibfnamefont {T.}~\bibnamefont {Kimura}},\ }\href
  {\doibase 10.1021/acs.inorgchem.5b02622} {\bibfield  {journal} {\bibinfo
  {journal} {Inorganic Chemistry}\ }\textbf {\bibinfo {volume} {55}},\ \bibinfo
  {pages} {1002} (\bibinfo {year} {2016}{\natexlab{b}})}\BibitemShut {NoStop}%
\bibitem [{\citenamefont {Babkevich}\ \emph {et~al.}(2017)\citenamefont
  {Babkevich}, \citenamefont {Testa}, \citenamefont {Kimura}, \citenamefont
  {Kimura}, \citenamefont {Tucker}, \citenamefont {Roessli},\ and\
  \citenamefont {R{\o}nnow}}]{babkevich2017magnetic}%
  \BibitemOpen
  \bibfield  {author} {\bibinfo {author} {\bibfnamefont {P.}~\bibnamefont
  {Babkevich}}, \bibinfo {author} {\bibfnamefont {L.}~\bibnamefont {Testa}},
  \bibinfo {author} {\bibfnamefont {K.}~\bibnamefont {Kimura}}, \bibinfo
  {author} {\bibfnamefont {T.}~\bibnamefont {Kimura}}, \bibinfo {author}
  {\bibfnamefont {G.}~\bibnamefont {Tucker}}, \bibinfo {author} {\bibfnamefont
  {B.}~\bibnamefont {Roessli}}, \ and\ \bibinfo {author} {\bibfnamefont
  {H.}~\bibnamefont {R{\o}nnow}},\ }\href {\doibase 10.1103/PhysRevB.96.214436}
  {\bibfield  {journal} {\bibinfo  {journal} {Phys. Rev. B}\ }\textbf {\bibinfo
  {volume} {96}},\ \bibinfo {pages} {214436} (\bibinfo {year}
  {2017})}\BibitemShut {NoStop}%
\bibitem [{\citenamefont {Kimura}\ \emph {et~al.}(2018)\citenamefont {Kimura},
  \citenamefont {Toyoda}, \citenamefont {Babkevich}, \citenamefont {Yamauchi},
  \citenamefont {Sera}, \citenamefont {Nassif}, \citenamefont {R{\o}nnow},\
  and\ \citenamefont {Kimura}}]{kimura2018cation}%
  \BibitemOpen
  \bibfield  {author} {\bibinfo {author} {\bibfnamefont {K.}~\bibnamefont
  {Kimura}}, \bibinfo {author} {\bibfnamefont {M.}~\bibnamefont {Toyoda}},
  \bibinfo {author} {\bibfnamefont {P.}~\bibnamefont {Babkevich}}, \bibinfo
  {author} {\bibfnamefont {K.}~\bibnamefont {Yamauchi}}, \bibinfo {author}
  {\bibfnamefont {M.}~\bibnamefont {Sera}}, \bibinfo {author} {\bibfnamefont
  {V.}~\bibnamefont {Nassif}}, \bibinfo {author} {\bibfnamefont
  {H.}~\bibnamefont {R{\o}nnow}}, \ and\ \bibinfo {author} {\bibfnamefont
  {T.}~\bibnamefont {Kimura}},\ }\href {\doibase 10.1103/PhysRevB.97.134418}
  {\bibfield  {journal} {\bibinfo  {journal} {Phys. Rev. B}\ }\textbf {\bibinfo
  {volume} {97}},\ \bibinfo {pages} {134418} (\bibinfo {year}
  {2018})}\BibitemShut {NoStop}%
\bibitem [{\citenamefont {Islam}\ \emph {et~al.}(2018)\citenamefont {Islam},
  \citenamefont {Ranjith}, \citenamefont {Baenitz}, \citenamefont {Skourski},
  \citenamefont {Tsirlin},\ and\ \citenamefont {Nath}}]{SS_Islam_SCTPO}%
  \BibitemOpen
  \bibfield  {author} {\bibinfo {author} {\bibfnamefont {S.~S.}\ \bibnamefont
  {Islam}}, \bibinfo {author} {\bibfnamefont {K.~M.}\ \bibnamefont {Ranjith}},
  \bibinfo {author} {\bibfnamefont {M.}~\bibnamefont {Baenitz}}, \bibinfo
  {author} {\bibfnamefont {Y.}~\bibnamefont {Skourski}}, \bibinfo {author}
  {\bibfnamefont {A.~A.}\ \bibnamefont {Tsirlin}}, \ and\ \bibinfo {author}
  {\bibfnamefont {R.}~\bibnamefont {Nath}},\ }\href {\doibase
  10.1103/PhysRevB.97.174432} {\bibfield  {journal} {\bibinfo  {journal} {Phys.
  Rev. B}\ }\textbf {\bibinfo {volume} {97}},\ \bibinfo {pages} {174432}
  (\bibinfo {year} {2018})}\BibitemShut {NoStop}%
\bibitem [{\citenamefont {Kato}\ \emph {et~al.}(2019)\citenamefont {Kato},
  \citenamefont {Kimura}, \citenamefont {Miyake}, \citenamefont {Tokunaga},
  \citenamefont {Matsuo}, \citenamefont {Kindo}, \citenamefont {Akaki},
  \citenamefont {Hagiwara}, \citenamefont {Kimura}, \citenamefont {Kimura},\
  and\ \citenamefont {Motome}}]{Y.Kato_2019_PRB}%
  \BibitemOpen
  \bibfield  {author} {\bibinfo {author} {\bibfnamefont {Y.}~\bibnamefont
  {Kato}}, \bibinfo {author} {\bibfnamefont {K.}~\bibnamefont {Kimura}},
  \bibinfo {author} {\bibfnamefont {A.}~\bibnamefont {Miyake}}, \bibinfo
  {author} {\bibfnamefont {M.}~\bibnamefont {Tokunaga}}, \bibinfo {author}
  {\bibfnamefont {A.}~\bibnamefont {Matsuo}}, \bibinfo {author} {\bibfnamefont
  {K.}~\bibnamefont {Kindo}}, \bibinfo {author} {\bibfnamefont
  {M.}~\bibnamefont {Akaki}}, \bibinfo {author} {\bibfnamefont
  {M.}~\bibnamefont {Hagiwara}}, \bibinfo {author} {\bibfnamefont
  {S.}~\bibnamefont {Kimura}}, \bibinfo {author} {\bibfnamefont
  {T.}~\bibnamefont {Kimura}}, \ and\ \bibinfo {author} {\bibfnamefont
  {Y.}~\bibnamefont {Motome}},\ }\href {\doibase 10.1103/PhysRevB.99.024415}
  {\bibfield  {journal} {\bibinfo  {journal} {Phys. Rev. B}\ }\textbf {\bibinfo
  {volume} {99}},\ \bibinfo {pages} {024415} (\bibinfo {year}
  {2019})}\BibitemShut {NoStop}%
\bibitem [{\citenamefont {Rodríguez-Carvajal}()}]{CARVAJAL_fullprof}%
  \BibitemOpen
  \bibfield  {author} {\bibinfo {author} {\bibfnamefont {J.}~\bibnamefont
  {Rodríguez-Carvajal}},\ }\href@noop {} {\bibinfo  {journal} {FullProf: A
  Program for Rietveld Refinement and Profile Matching Analysis of Complex
  Powder Diffraction Pattern (ILL, unpublished)}\ }\BibitemShut {NoStop}%
\bibitem [{\citenamefont {Haraldsen}\ \emph {et~al.}(2005)\citenamefont
  {Haraldsen}, \citenamefont {Barnes},\ and\ \citenamefont
  {Musfeldt}}]{Formula_Rectangular_Tetramer}%
  \BibitemOpen
\bibfield  {journal} {  }\bibfield  {author} {\bibinfo {author} {\bibfnamefont
  {J.}~\bibnamefont {Haraldsen}}, \bibinfo {author} {\bibfnamefont
  {T.}~\bibnamefont {Barnes}}, \ and\ \bibinfo {author} {\bibfnamefont
  {J.}~\bibnamefont {Musfeldt}},\ }\href {\doibase 10.1103/PhysRevB.71.064403}
  {\bibfield  {journal} {\bibinfo  {journal} {Physical Review B}\ }\textbf
  {\bibinfo {volume} {71}},\ \bibinfo {pages} {064403} (\bibinfo {year}
  {2005})}\BibitemShut {NoStop}%
\bibitem [{\citenamefont {Kato}\ \emph {et~al.}(2017)\citenamefont {Kato},
  \citenamefont {Kimura}, \citenamefont {Miyake}, \citenamefont {Tokunaga},
  \citenamefont {Matsuo}, \citenamefont {Kindo}, \citenamefont {Akaki},
  \citenamefont {Hagiwara}, \citenamefont {Sera}, \citenamefont {Kimura},\ and\
  \citenamefont {Motome}}]{kato2017_theory_fit}%
  \BibitemOpen
  \bibfield  {author} {\bibinfo {author} {\bibfnamefont {Y.}~\bibnamefont
  {Kato}}, \bibinfo {author} {\bibfnamefont {K.}~\bibnamefont {Kimura}},
  \bibinfo {author} {\bibfnamefont {A.}~\bibnamefont {Miyake}}, \bibinfo
  {author} {\bibfnamefont {M.}~\bibnamefont {Tokunaga}}, \bibinfo {author}
  {\bibfnamefont {A.}~\bibnamefont {Matsuo}}, \bibinfo {author} {\bibfnamefont
  {K.}~\bibnamefont {Kindo}}, \bibinfo {author} {\bibfnamefont
  {M.}~\bibnamefont {Akaki}}, \bibinfo {author} {\bibfnamefont
  {M.}~\bibnamefont {Hagiwara}}, \bibinfo {author} {\bibfnamefont
  {M.}~\bibnamefont {Sera}}, \bibinfo {author} {\bibfnamefont {T.}~\bibnamefont
  {Kimura}}, \ and\ \bibinfo {author} {\bibfnamefont {Y.}~\bibnamefont
  {Motome}},\ }\href {\doibase 10.1103/PhysRevLett.118.107601} {\bibfield
  {journal} {\bibinfo  {journal} {Phys. Rev. Lett.}\ }\textbf {\bibinfo
  {volume} {118}},\ \bibinfo {pages} {107601} (\bibinfo {year}
  {2017})}\BibitemShut {NoStop}%
\bibitem [{\citenamefont {E.S.R.}(2012)}]{gopal2012specific}%
  \BibitemOpen
  \bibfield  {author} {\bibinfo {author} {\bibfnamefont {G.}~\bibnamefont
  {E.S.R.}},\ }\href@noop {} {\emph {\bibinfo {title} {Specific heats at low
  temperatures}}}\ (\bibinfo  {publisher} {Springer Science \& Business
  Media},\ \bibinfo {year} {2012})\BibitemShut {NoStop}%
\bibitem [{\citenamefont {Ashcroft}\ and\ \citenamefont
  {Mermin}(1976)}]{ashcroft1976solid}%
  \BibitemOpen
  \bibfield  {author} {\bibinfo {author} {\bibfnamefont {N.}~\bibnamefont
  {Ashcroft}}\ and\ \bibinfo {author} {\bibfnamefont {N.}~\bibnamefont
  {Mermin}},\ }\href {https://books.google.co.in/books?id=1C9HAQAAIAAJ} {\emph
  {\bibinfo {title} {Solid State Physics}}},\ HRW international editions\
  (\bibinfo  {publisher} {Holt, Rinehart and Winston},\ \bibinfo {year}
  {1976})\BibitemShut {NoStop}%
\bibitem [{\citenamefont {Joshua}(1998)}]{joshua1998magnon}%
  \BibitemOpen
  \bibfield  {author} {\bibinfo {author} {\bibfnamefont {S.}~\bibnamefont
  {Joshua}},\ }\href {\doibase https://doi.org/10.1016/S0378-4371(98)00370-7}
  {\bibfield  {journal} {\bibinfo  {journal} {Physica A: Statistical Mechanics
  and its Applications}\ }\textbf {\bibinfo {volume} {261}},\ \bibinfo {pages}
  {135} (\bibinfo {year} {1998})}\BibitemShut {NoStop}%
\bibitem [{\citenamefont {Pincus}\ and\ \citenamefont
  {Winter}(1961)}]{Pincuss-Winter}%
  \BibitemOpen
  \bibfield  {author} {\bibinfo {author} {\bibfnamefont {P.}~\bibnamefont
  {Pincus}}\ and\ \bibinfo {author} {\bibfnamefont {J.}~\bibnamefont
  {Winter}},\ }\href {\doibase 10.1103/PhysRevLett.7.269} {\bibfield  {journal}
  {\bibinfo  {journal} {Phys. Rev. Lett.}\ }\textbf {\bibinfo {volume} {7}},\
  \bibinfo {pages} {269} (\bibinfo {year} {1961})}\BibitemShut {NoStop}%
\bibitem [{\citenamefont {Moriya}(1956)}]{moriya1956nuclear}%
  \BibitemOpen
  \bibfield  {author} {\bibinfo {author} {\bibfnamefont {T.}~\bibnamefont
  {Moriya}},\ }\href {\doibase https://doi.org/10.1143/PTP.16.23} {\bibfield
  {journal} {\bibinfo  {journal} {Progress of Theoretical Physics}\ }\textbf
  {\bibinfo {volume} {16}},\ \bibinfo {pages} {641} (\bibinfo {year}
  {1956})}\BibitemShut {NoStop}%
\bibitem [{\citenamefont {Moriya}(1963)}]{moriya1963}%
  \BibitemOpen
  \bibfield  {author} {\bibinfo {author} {\bibfnamefont {T.}~\bibnamefont
  {Moriya}},\ }\href {\doibase 10.1143/JPSJ.18.516} {\bibfield  {journal}
  {\bibinfo  {journal} {Journal of the Physical Society of Japan}\ }\textbf
  {\bibinfo {volume} {18}},\ \bibinfo {pages} {516} (\bibinfo {year}
  {1963})}\BibitemShut {NoStop}%
\bibitem [{\citenamefont {Mahajan}\ \emph {et~al.}(1998)\citenamefont
  {Mahajan}, \citenamefont {Sala}, \citenamefont {Lee}, \citenamefont {Borsa},
  \citenamefont {Kondo},\ and\ \citenamefont {Johnston}}]{Mahajan_Li2VO4}%
  \BibitemOpen
  \bibfield  {author} {\bibinfo {author} {\bibfnamefont {A.~V.}\ \bibnamefont
  {Mahajan}}, \bibinfo {author} {\bibfnamefont {R.}~\bibnamefont {Sala}},
  \bibinfo {author} {\bibfnamefont {E.}~\bibnamefont {Lee}}, \bibinfo {author}
  {\bibfnamefont {F.}~\bibnamefont {Borsa}}, \bibinfo {author} {\bibfnamefont
  {S.}~\bibnamefont {Kondo}}, \ and\ \bibinfo {author} {\bibfnamefont {D.~C.}\
  \bibnamefont {Johnston}},\ }\href {\doibase 10.1103/PhysRevB.57.8890}
  {\bibfield  {journal} {\bibinfo  {journal} {Phys. Rev. B}\ }\textbf {\bibinfo
  {volume} {57}},\ \bibinfo {pages} {8890} (\bibinfo {year}
  {1998})}\BibitemShut {NoStop}%
\bibitem [{\citenamefont {Kaplan}\ \emph {et~al.}(1966)\citenamefont {Kaplan},
  \citenamefont {Loudon}, \citenamefont {Jaccarino}, \citenamefont
  {Guggenheim}, \citenamefont {Beeman},\ and\ \citenamefont
  {Pincus}}]{N_Kaplan}%
  \BibitemOpen
  \bibfield  {author} {\bibinfo {author} {\bibfnamefont {N.}~\bibnamefont
  {Kaplan}}, \bibinfo {author} {\bibfnamefont {R.}~\bibnamefont {Loudon}},
  \bibinfo {author} {\bibfnamefont {V.}~\bibnamefont {Jaccarino}}, \bibinfo
  {author} {\bibfnamefont {H.~J.}\ \bibnamefont {Guggenheim}}, \bibinfo
  {author} {\bibfnamefont {D.}~\bibnamefont {Beeman}}, \ and\ \bibinfo {author}
  {\bibfnamefont {P.~A.}\ \bibnamefont {Pincus}},\ }\href {\doibase
  10.1103/PhysRevLett.17.357} {\bibfield  {journal} {\bibinfo  {journal} {Phys.
  Rev. Lett.}\ }\textbf {\bibinfo {volume} {17}},\ \bibinfo {pages} {357}
  (\bibinfo {year} {1966})}\BibitemShut {NoStop}%
\bibitem [{\citenamefont {Beeman}\ and\ \citenamefont
  {Pincus}(1968)}]{D_Beeman}%
  \BibitemOpen
  \bibfield  {author} {\bibinfo {author} {\bibfnamefont {D.}~\bibnamefont
  {Beeman}}\ and\ \bibinfo {author} {\bibfnamefont {P.}~\bibnamefont
  {Pincus}},\ }\href {\doibase 10.1103/PhysRev.166.359} {\bibfield  {journal}
  {\bibinfo  {journal} {Phys. Rev.}\ }\textbf {\bibinfo {volume} {166}},\
  \bibinfo {pages} {359} (\bibinfo {year} {1968})}\BibitemShut {NoStop}%
\end{thebibliography}%

\end{document}